\documentclass[fleqn,10pt]{SelfArx}
\usepackage{adjustbox}
\usepackage{adjustbox}
\usepackage{amsmath}
\usepackage{amssymb}
\usepackage{float}
\usepackage[T1]{fontenc}
\usepackage{graphicx}
\usepackage{hhline}
\usepackage[utf8]{inputenc}
\usepackage{mathtools}
\usepackage{multicol}
\usepackage{multirow}
\usepackage{wasysym}
\usepackage{cite}
\usepackage{algorithmic}
\usepackage{textcomp}
\usepackage{xcolor}
\usepackage{amsmath}
\usepackage{graphicx}
\usepackage[hidelinks, colorlinks=true, allcolors=blue]{hyperref}

\usepackage{lipsum} % Required to insert dummy text. To be removed otherwise
\definecolor{color1}{RGB}{0,0,90} % Color of the article title and sections
\definecolor{color2}{RGB}{0,20,20} % Color of the boxes behind the abstract and headings

\usepackage{hyperref} % Required for hyperlinks
\hypersetup{hidelinks,colorlinks,breaklinks=true,urlcolor=color2,citecolor=color1,linkcolor=color1,bookmarksopen=false,pdftitle={Development of user’s trend in volunteer-based networks over time},pdfauthor={Author}}

\JournalInfo{Draft Version} % Journal information
\Archive{}

\PaperTitle{Analyzing Key Users' behavior trends in Volunteer-Based Networks} % Article title

\Authors{Nofar Piterman\textsuperscript{1}*, Tamar Makov\textsuperscript{2}, and Michael Fire\textsuperscript{1}} % Authors
\affiliation{\textsuperscript{1}\textit{Department of Software and Information Systems Engineering, Ben-Gurion University, Beersheba, Israel}} % Author affiliation

\affiliation{\textsuperscript{2}\textit{Department of Management, Guilford Glazer Faculty of Business and Management, Ben-Gurion University, Beersheba, Israel}} % Author affiliation Corresponding author
 % Keywords - if you don't want any simply remove all the text between the curly brackets
\newcommand{\keywordname}{Keywords} % Defines the keywords heading name

\Abstract{
Online social networks usage has increased significantly in the last decade and continues to grow in popularity. Multiple social platforms use volunteers as a central component. The behavior of volunteers in volunteer-based networks has been studied extensively in recent years. Here, we explore the development of volunteer-based social networks, primarily focusing on their key users’ behaviors and activities. We developed two novel algorithms: the first reveals key user behavior patterns over time; the second utilizes machine learning methods to generate a forecasting model that can predict the future behavior of key users, including whether they will remain active donors or change their behavior to become mainly recipients, and vice-versa. These algorithms allowed us to analyze the factors that significantly influence behavior predictions.

To evaluate our algorithms, we utilized data from over 2.4 million users on a peer-to-peer food-sharing online platform. Using our algorithm, we identified four main types of key user behavior patterns that occur over time. Moreover,
we succeeded in forecasting future active donor key users and predicting the key users that would change their behavior to donors, with an accuracy of up to 89.6\%. These findings provide valuable insights into the behavior of key users in volunteer-based social networks and pave the way for more effective communities-building in the future, while using the potential of machine learning for this goal.
}

\begin{document}
\maketitle % Print the title and abstract box
\flushbottom % Makes all text pages the same height
%\tableofcontents % Print the contents section

\thispagestyle{empty} %Removes page numbering from the first page

\section{Introduction}
\label{section:Introduction}
\addcontentsline{}{}{}
\addcontentsline{toc}{section}{Introduction}
Social media platforms have become an integral part of our lives, with 4.5 billion people worldwide using them~\cite{socialMediaUsers}. Several social networks use volunteers as a central network component~\cite{OLIOex, dosono2019moderation, jhaver2019human}. The recent years have witnessed a lively scholarly interest in the study of volunteers within volunteer-based networks, with many exploring their behaviors and whether these can be predicted accurately~\cite{song2015multiple, song2016volunteerism}. 

Volunteers are a crucial element of many networks, where they are responsible for a range of tasks, from managing communities, to creating and maintaining content. For instance, volunteer moderators on Reddit\footnote{\url{http://www.reddit.com}} work to ensure that the platform remains active~\cite{dosono2019moderation}, while Wikipedia\footnote{\url{http://www.wikipedia.org}} is maintained by thousands of volunteers who contribute to its community structure~\cite{baytiyeh2010volunteers}. OLIO,\footnote{\url{http://www.olioex.com}} on the other hand, is a real-world and online network where its volunteers are known as "food waste heroes" and depended on for their participation in food-sharing transactions~\cite{dosono2019moderation, baytiyeh2010volunteers, OLIOex}.

Operating a volunteer-based network can present several challenges for companies. These include the challenge of identifying and recruiting volunteers~\cite{song2015multiple, song2016volunteerism}. Additionally, companies such as Reddit must manage and mitigate the potential for volunteer exhaustion, while also ensuring efficient volunteer engagement~\cite{dosono2019moderation}.

%Volunteers are important elements of the network and are involved in many activities. According to network science, in many networks, users (referred to as hubs) which have a relatively high number of connections have a more substantial effect on the network than others. It is manifested in their contribution to the development of the network. Therefore, volunteers are a main component of the network and its dynamics. In many cases, these types of users are in high probability to attract much more new connections in the future than regular nodes~\cite{brintrup2015supply}. 

This study explores the development of real-world volunteer-based networks over time-based network users’ giving and taking transactions. These types of networks, such as OLIO and Reddit,\footnote{\url{http://www.reddit.com}} are primarily built on volunteers who have key roles in the network’s activities and dynamics.

In this study, we developed two novel methods. The first uncovers different key users’ behavior patterns over time. The second enables the forecasting of the future behaviors of key users by using supervised learning algorithms. It can predict whether a key user will either be a primarily active donor or change their behavior and become mainly a recipient. This method makes it possible to predict which key users will become active donors and to forecast which of those active donors will decrease their donation activities.

Our first method (see Section~\ref{subsection:Analyze key users of the network}), identifies the different patterns of key users’ behavior. This includes several main steps: first, working with given data on transactions between users, we preprocess this by filtering the users who use the network less than the minimal amount of time for analyzing (detailed in Section~\ref{section:METHODS}) and have a relatively low number of transactions. Next, we construct a network based on the transactions of the filtered users, where each node represents a user, and each edge between two nodes represents transactions between two users.

Subsequently, we analyze the key users in the network by defining an innovative measure, the \textit{Donors ratio (DR)} for user behavior. The DR measure is based on the user’s network transactions of giving and taking. User behavior is shown by this new measure- whether the user is primarily a donor or a beneficiary. We use this measure to calculate a series of values representing each key user’s behavior over time and analyze the changes, such as changes from active roles to beneficiary ones, or vice versa. Based on this calculated time-series measure, a time-series algorithm clusters key users according to their behaviors, resulting in groups that demonstrate different behavior patterns.
%Using this series as an input to a time-series clustering algorithm, we cluster the users into groups based on similar donation patterns. 
%We calculate this measure for the users at a specific time in the life cycle of their use in the network. 

Our second method (see Section~\ref{subsection:Users’ trends prediction}) predicts the key user’s future behavior based on their association with one of the clusters. We use features extracted from the stored social network’s raw data regarding the key user parameters, in addition to network features based on graph theory (see Sections~\ref{subsubsection:Features Extraction} and~\ref{subsection:Evaluate}). We use state-of-the-art machine learning (ML) prediction models, such as RandomForest~\cite{biau2016random} and XGBoost~\cite{yacouby2020probabilistic}to predict which group the user will belong to, based on the similarity of their behavior to other group members.

To test and evaluate our models, we utilized data from OLIO, a peer-to-peer (P2P) food-sharing platform that aims to diminish global food waste~\cite{pasqualemarcellofalcone2017bringing}. 
%OLIO connects private people and local businesses in the real world so surplus food can be shared, not thrown away
OLIO provided us with its dataset, which was collected over the course of 40 months, containing over 2.48 million users and 2.65 million items worldwide, with an average of 600k user transactions per month. We focused only on data from different geographical locations within the United Kingdom (UK).

%OLIO's users open the app, add a photo of the item, description, when and where it is available for pick up and wait for the other users to respond. To access an item, the users browse the listing near them, request the items they want, and then both users, the taker, and the giver arrange a pick-up via private messaging.

OLIO’s network uses predefined key users, which are super donor users in the network: the \textit{food waste heroes}. Food waste hero (hereafter referred to as “heroes”) is a title for official OLIO volunteers who collect surplus food from local businesses (e.g., supermarkets or delis), saving it from going to waste by redistributing it to their neighbors, who pick up the food. The heroes significantly affect the amount of surplus food redistributed via OLIO: they are the primary sources of supply on the platform~\cite{makov2020social}.

%Despite the fact that there are fewer food waste heroes than regular users (about 20\% in the largest sub-network), food waste heroes' transactions account for about 57\% of network transactions in the largest sub-networks. 

To assess the performance of our models, we carried out an empirical study of the key users of OLIO. Our aim was to gain insight into their behavior patterns over time and their overall impact on the network. Specifically, we utilized the first method, which examined their listing and pickup behaviors over time, calculating the percentage of items listed by each hero by dividing this number by the total number of items they listed and picked up.

\begin{figure*}[ht]
    \centering
    \includegraphics[width=0.9\linewidth]{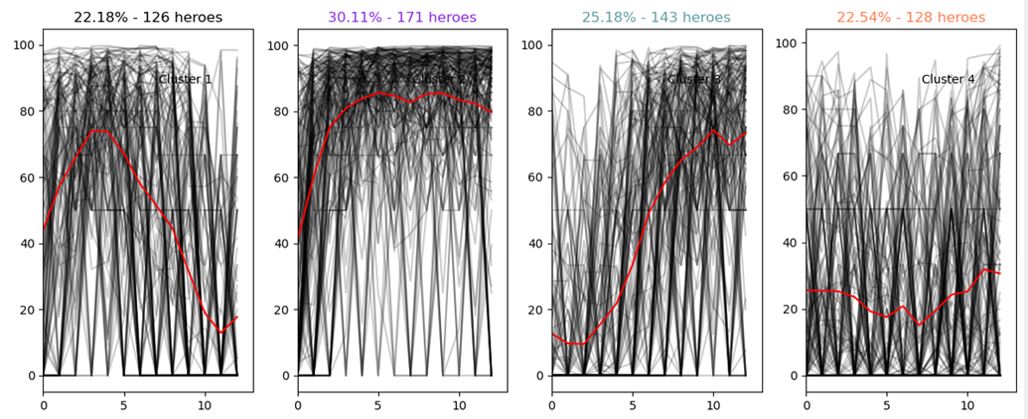}
    \caption{Four clusters of listing percent of users}
    \label{figure:12}
\end{figure*}

Using our algorithm, we identified four main types of groups:

\begin {enumerate}

\item \textit{Future Passive Donors} - users whose initial percentage of listing items is \textit{high and then decreases}.

\item \textit{Stable Active Donors} - users whose initial percentage of listing items is \textit{high and remains stable}.

\item \textit{Future Active Donors} - users whose initial percentage of listing items is \textit{low and then increases}.

\item \textit{Stable Passive Donors} - users whose initial percentage of listing items is\textit{ low and remains stable}. 

\end{enumerate}
%The first group is users which reduces their listing activity and their food distribution. The second and fourth groups are volunteers whose behavior is stable over time and the third group is users who mainly pick up items and change their behavior toward donors over time. 

To test and evaluate our second method, we created two binary prediction models which anticipate the future behavior of each key user based on the behavior of similar users. The first relates to the case of key users with low initial listing items percentage (starting as a "passive donor") and predict whether the key user stays stable as a "passive donor" or changes their behavior to an "active donor". The second relates to key users with a high initial listing items percentage (starting as an "active donor"), where we tested and evaluated our proposed methods on the entire UK network and its largest communities. In our experiment on the UK’s largest community with key users who have low initial listing items percentage, we got an accuracy score of 89.6\%. 

This study's main contributions are twofold:
\begin{itemize}

\item We present a novel algorithm for analyzing volunteer-based networks based on giving and taking transactions, in order to identify the network’s key users and explore various types of users that display different trends of behaviors over time.

\item We present a forecast of key users’ behavioral changes in their network usage over time, based on machine learning algorithms. By predicting user behavior, organizations can distribute resources correctly and efficiently between different key users, which will positively benefit the network’s development.

\end{itemize}

The rest of this paper is organized as follows: Section~\ref{section:RELATED WORK} provides an overview of related studies in the fields of volunteer-based networks, OLIO, network analysis, and time-series clustering. Section~\ref{section:METHODS} describes our methods- identification of different patterns of key users’ behaviors, and user behavior prediction. Section~\ref{section:EXPERIMENTS} details the experiments of our proposed method and then Section~\ref{section:RESULTS} sets out our results. In Section~\ref{section:DISCUSSION}, we discuss the obtained results and then finally, Section~\ref{section:CONCLUSION} presents our conclusions and offer future research directions.

\section{RELATED WORK}
\label{section:RELATED WORK}
Extensive research has been conducted in recent years on the examination of volunteers within volunteer-based networks and the forecasting of their behavior~\cite{song2015multiple, song2016volunteerism, baytiyeh2010volunteers}.
In this section, offer an overview of the studies related to our research on two main topics. Firstly, in Section~\ref{subsection:volunteer-based networks and OLIO}, we provide an overview of volunteer-based networks, emphasizing the previous studies that utilized OLIO’s network dataset. Secondly, Section~\ref{subsection:Network analysis} reviews related critical works in network analysis and community detection methods. Lastly, Section~\ref{subsection:Time-series Clustering} explores relevant time-series clustering methods and prediction models.

\subsection{Volunteer-based networks and OLIO}
\label{subsection:volunteer-based networks and OLIO}
The scope of volunteer-based networks has grown in recent years. Song et al.’s 2015 study highlights the significance of volunteerism and the challenges of accurately predicting volunteer tendencies.

In 2016, the follow-up study of Song et al.~\cite{song2016volunteerism} discussed the prediction of volunteerism tendencies through the harvesting of multiple social networks. It recognizes the importance of volunteerism and the challenges associated with accurately predicting volunteer behavior. The authors propose a methodology that involves gathering data from various social networks to improve prediction accuracy. They analyze user profiles, social relationships, and activity patterns to develop predictive models.

Wikipedia is another successful network that mainly relies on volunteers who form the foundation of its operations and contribute significantly to its activities~\cite{baytiyeh2010volunteers}. Baytiyeh et al.'s~\cite{baytiyeh2010volunteers} explored the importance of Wikipedia’s volunteers and highlighted the significance of the community behind the platform. They explored the reasons why volunteers play a crucial role in Wikipedia’s success and sustainability, and indicated the vital role of Wikipedia’s volunteer community in its development and maintenance as a reliable and comprehensive source of information.

Another well-known volunteer-based network is Reddit, which was the subject of Dosono et al.’s study~\cite{dosono2019moderation}. They employed a qualitative research approach to examine the role of moderators—who are volunteers and key users—in upholding the community and creating a conducive environment for discussions. Their findings demonstrated the significant impact of these key users on the network’s sustainability and progress.

Over the last decades, there has been a remarkable increase in the popularity of sharing economies. For example, the Freecycle network, which was founded in May 2003, is a volunteer-based network whose members make transactions by gifting objects to strangers~\cite{aptekar2016gifts}. This network was studied by Aptekar~\cite{aptekar2016gifts} , who highlights the role of trust, reciprocity, social norms in the exchange of items and the sense of community among participants.

Volunteer-based networks can emerge across diverse fields. One such example is the OLIO app, which was founded in late 2015 by Tessa Clarke and Saasha Celestial-One~\cite{OLIOex}. The app operates in the field of technological progress and modernization, working to reduce food waste through its network of volunteer sharing. Over the years, the data collected from food-sharing platforms, like OLIO, have been studied extensively~\cite{makov2020social,morone2019food,michelini2019uncovering,harvey2019food,gonzalezraya2021upscaling,nica2021identifying}. In 2017, Michelini et al.~\cite{michelini2017understanding} utilized 52 food-sharing platforms, including OLIO, to examine how digital opportunities redesign alternative distribution systems, and how online contexts impact their value. Their results indicate the existence of three main models for food-sharing platforms: first, the money-sharing B2C model, whose primary goal is to reduce food waste while creating profits. Second, is charity sharing: a model in which food is donated to non-profit associations. Thirdly, a P2P model uses a community-based process in which food is shared. Additionally, Michelini et al. [18] indicated three main features which contributed to a better understanding of the differences between the models: delivery models, the type of donor/beneficiary, and type of transaction. Lastly, they specified that those features are more significant than technological characteristics and the geographical area covered.

In 2019, Michelini et al.~\cite{michelini2019uncovering} investigated the potential impact of food-sharing platform business models. They focused on two leading platforms: OLIO and Too Good To Go.\footnote{\url{http://www.toogoodtogo.org}} They showed that an extensive community in terms of size and high volunteer involvement is necessary for successful network development in both cases~\cite{michelini2019uncovering}. In the same year, Harvey et al.~\cite{harvey2019food} analyzed how OLIO has changed the conventional food supply chain through mobile applications, exploring food-sharing, redistribution, and waste reduction. The study examined whether most OLIO users were donors or recipients, and whether their behavior patterns are consistent, examining the relationships and transactions between users over time. They discovered that there is a qualitative split in the number of users that primarily give or take. The study does not reveal how users of OLIO behave over an extended period, but does provide compelling evidence for various enacted behavioral roles. Additionally, Harvey et al. point out that “the role of volunteers and commercial donors inevitably has a strong impact on the anatomy of the network”~\cite{harvey2019food}. 

Nica-Avram et al.~\cite{nica2021identifying} analyzed OLIO’s network to identify food insecurity in 2020. They found that most individuals used this platform to either donate or request food, but not both. They also identified that heroes played the most active role among users. In the same year, Makov et al.~\cite{makov2020social} analyzed data provided by OLIO between January 2018 and May 2020.  %They examined only food items. 
Their research included several parts: all listings were classified into food categories using a supervised deep-learning long short-term memory (LSTM) network; the results showed that of the 22,000 users who performed at least one transaction of food exchange, 12\% both gave and took at least one item, 26\% had only given, and 62\% only collected; and they saw that heroes were~\cite{OLIOex} more likely to engage with a larger number of users (27 on average compared to 2.5 for regular users)~\cite{makov2020social}. 

In 2021, a qualitative study by Federico Gonzalez Raya~\cite{gonzalezraya2021upscaling} sshowed that city residents could have varying roles regarding food-sharing within the city: they can be both consumers and producers; not just consumers. The study used interview transcripts and field observations, which were performed on three case studies: OLIO, Foodsharing,\footnote{foodsharing.de} and Reko.\footnote{localfoodnodes.org/en/node/reko-stockholm-liljeholmen} Furthermore, they showed that a user’s role can evolve and change from a passive consumer to an active consumer in the city’s food-sharing network k~\cite{gonzalezraya2021upscaling}.

Recently, Makov et al.~\cite{makov2023digital} studied food insecurity during the COVID-19 era, using OLIO’s network data. In order to compare this time period to its prior, they looked at the ratio between users who are providers to users who are collecting over time, on a weekly basis. They define a provider user as a user who gives one or more items in a week. Likewise, they define a collector user as someone who collects one or more items in a week.

\subsection{Network analysis}
\label{subsection:Network analysis}
Previous work has been done in social network analysis regarding types of networks, their topology, and related algorithms, some of which helped build the ground for our study’s intervention. 

Our study utilized network algorithms with different network measures (see Section~\ref{section:METHODS}) to test the similarity between users in the networks to predict users’ behaviors. Network measures can be calculated from two aspects: the node and the network~\cite{kim2010structural}. According to Kim et al.~\cite{kim2010structural}, it is necessary to address not only the network perspective, but also the node elements to get the entire perspective of the network analysis. Node-level metrics measure how an individual node is ingrained in a network from that particular node’s perspective. Kim et al.’s research focuses on three types of node-level metrics: degree, closeness, and betweenness centrality.

In 2018, Chakraborty et al.~\cite{chakraborty2018application} presented two approaches for network analysis, the first being the “overall network”, which is based on the ties and relationships between all users or other objects in the network. The second approach focuses on individual users and their close neighbors in the network. Additionally, they describe eight different measures for networks, including: size, density, connectedness, diameter and average path length, clustering, centrality, and degree distributions. Finally, they compared different networks based on those measures~\cite{chakraborty2018application}.

Social networks are based on transactions between different users in the network. In this study, we tested our methods on the volunteer-based social network OLIO, which has no limit on the number of transactions that can be made by each user. This is reflected in the heroes’ activity: a previous study showed that heroes are more likely to engage with many users and thus do more transactions than regular users~\cite{makov2020social}.

In addition, in this study, we utilize a variety of network measure (see Section~\ref{subsubsection:Features Extraction}). Our proposed method shares similarities with Harvey et al.’s study~\cite{harvey2019food}, which focused on specific food chains and food-sharing networks, and also to Nica-Avram et al.~\cite{nicaavram2021identifying}, where the relationship between food-sharing and deprivation by analyzing OLIO’s network was explored.

Several studies have analyzed the OLIO network to test different cases. Harvey et al.~\cite{michelini2019uncovering} core method was exploratory social network analysis of food-sharing mobile applications undertaken in partnership with OLIO. Their method was divided into two main parts. The first used basic network measures to assess donor-recipient reciprocity and balance, and the second calculated the importance of a specific node in the social network complex. Network measures were used to determine interdependence based on different centrality measures, where centrality was measured according to a node’s degree, closeness, betweenness, eigenvalue, or power measures~\cite{bloch2016Centrality,Landherr2010review}. Their method utilized those measures on OLIO’s data to get statistical information about donors and recipients and understand how individual users affect the network behavior~\cite{harvey2019food}.

In their study, Nica-Avram et al.~\cite{nicaavram2021identifying} examined the food-sharing behaviors of OLIO users in the UK. Their main methodology was based on OLIO’s observed and inferred data. One of the approaches used was the analysis of network typologies. Their findings revealed that the majority of users utilized OLIO to either request or donate food, while only a small proportion of users engaged in both roles. These results are consistent with the findings of Makov et al.~\cite{makov2020social}, who discovered that the largest group of users consisted of individuals who collected items, followed by the second largest group, which comprised users who listed items. Only a small number of users participated in both activities. These findings imply a significant polarization between different users’ roles, where the ratio between the number of their listed items and the number of all their transactions will be extremely high or extremely low. However, it is rarely in the middle.

One of this study’s main goals is to predict the development and the success or failure of a food-sharing network. A similar goal motivated in a recent study by Mazzucchelli et al.~\cite{mazzucchelli2021how}], who set out to understand how and to what extent an online food-sharing network is successful. They identified the main causes of higher consumer engagement within these online platforms. To achieve this, they analyzed data obtained from 455 users of the OLIO platform. Mazzucchelli et al. applied a multivariate regression analysis, with the dependent variable being the consumer behavioral response related to users’ utilization of the OLIO mobile application, as either food donors or recipients. The study aims to assist different online food-sharing platforms in creating and implementing a successful network that motivates them to join and be a part of a community.

Another key social network method used in our study is community detection~\cite{bedi2016community,ponveni2021review}. Social networks are characterized by small communities with similar properties (sometimes based on geographical areas). Grouping users into communities is useful in identifying the connection among the nodes and finding characteristics by using similar nodes in the community~\cite{ponveni2021review}, a common research area in collaboration networks, a set of like-minded users for marketing and recommendations~\cite{bedi2016community}, and analyzing the network development.

The community detection algorithms’ goal is to discover the groups formed in the network. In many applications where group decisions are made, identifying communities can be helpful. Entities in one community will be closer to each other and interact with each other more frequently than entities in different communities. The closeness between entities in a group can be calculated with distance or similarity measures between entities~\cite{bedi2016community}. Accordingly, based on previous studies~\cite{jain2019discover, sanchez2016twitter}, we use a common community detection algorithm—the Louvain algorithm~\cite{ghosh2018distributed}, which is an agglomerative approach, to analyze and extract communities from large networks.

\subsection{Time-series Clustering}
\label{subsection:Time-series Clustering}
There has been a growth in the use of time-series data measured at regular intervals of time in many fields in recent years. Amongst this data, there are networks and systems~\cite{ali2019clustering}, in addition to chronological observations over sequential time ~\cite{fu2011a}.

This study used the user’s transactions as time-series data for pattern discovery, which is the most common mining task related to time-series data~\cite{fu2011a}. We analyzed users’ behavior related to their number of listings and the number of pickups of items each week over one year. Our goal is to discover future trends in user activity and classify the different types of users. Therefore, we use time-series clustering algorithms to cluster similar users by their behavior~\cite{ali2019clustering}. One of the most common ways to address this issue is by using distance-based clustering approaches. The choice of similarity measures seriously influences the quality of mining techniques.

Our research is based on Ruiz et al. study~\cite{ruiz2020} that tested different clustering methods and Ali et al.~\cite{ali2019clustering} study, which described Euclidean and Dynamic Time Warping (DTW) distance measures. After testing both distance measures, we utilized the K-means algorithm~\cite{kanungo2000the} with Euclidean distance to cluster users by their behavior.

Ruiz et al.~\cite{ruiz2020} tested and compared several well-known time-series methods and several distance matrices on energy consumption data, to select the most appropriate model. They tested four different models: K-means, kmedoids, Hierarchical clustering, and Gaussian Mixtures ~\cite{ruiz2020}. They found that the most effective results were obtained by K-means and k-medoids, which showed similar results. The best distance metric for the K-means method was Squared Euclidean distance, outperforming other metrics.

Ali et al.~\cite{ali2019clustering} described two similar measures: Euclidean distance and DTW. The Euclidean distance between two timeseries is the square root of the sum of the squared differences, which is calculated by matching the corresponding points along the horizontal axis. On the other hand, the concept of the DTW method is to warp the series before computing the distance. Nevertheless, two temporal points with totally different local structures may not be fitted correctly by DTW. According to Ali et al., Euclidean Distance—a prevalent distance measure in the surveyed visual analytical papers—is, compared to other similarity measures, clear and straightforward.

\section{METHODS}
\label{section:METHODS}
Our goal is to learn about the development of OLIO’s network over time. In order to track this, we define two methods. We first analyze key users and uncover different behavior patterns. Second, we utilize these patterns to identify changes in key users’ behavior, or find unusual behavior that may interfere with improvement in the development of the user or network. In the following subsections, we detail those two methods.

\subsection{Analyze key users of the network}
\label{subsection:Analyze key users of the network}
We perform two main parts to analyze key users and their trends in behavior over time (see Figure~\ref{figure:2}). The first part focuses on detecting key users and calculating a behavior measure at fixed time intervals (see Section~\ref{subsection:Analyze key users of the network}). The second part focuses on uncovering different types of key users’ behavioral trends (see Section~\ref{subsection:Users’ trends prediction}). 

Given a dynamic social network, we can represent the network structure at time $t$ by a dynamic directed graph, $G^{t} := <V^{t},E^{t}>$, with a set of nodes $V^{t}$ which represent network users, and a set of edges $E^{t}$, which represent transactions between the users that occurred until time $t$.

To identify and analyze the key users of the network, we investigated the network in a predefined time period and carried out three steps. First, the key users need to be identified. Two options are available here: using predefined nodes or performing network exploration and node analysis to uncover the key users based on the network’s topology. An example for node analysis is that node will act as a key node if it meets the definition of hub.\footnote{A hub is defined as a node $v\in V$, with a number of edges entering and exiting a node, defined as the node’s degree (denoted $d^{t}(v)$), that holds $d^{t}(v)\gg d_{avg}^{t}$, where $d_{avg}^{t} =\frac{\sum_{v\in V}d^{t}(v)}{\vert V\vert }$ at time $t$. Namely, a node $v$ is defined as a hub if it has a higher degree than the average degree of the nodes in the network~\cite{zhou2008brief}.}
During our study, we evaluated our methods using the OLIO network with predefined list of key users which are the food waste heroes. These users were chosen because they are known to participate in a large number of transactions based on previous research~\cite{makov2020social,nica2021identifying}. Next, for each node representing a key user, we defined and calculated the Donors Ratio (referred to as DR) measure of the user’s transactions. We defined DR for a user $u\in Users$ at a specific time interval $\Delta t=[t_{1},t_{2}]$ as the ratio between the number of listing transactions of a user in interval time $\Delta t$ and the number of all his or her transactions (listing and pickup) in the same interval time $\Delta t$. Namely, for a user $u\in Users$, we define Donors Ratio as:
\[DR(u,\Delta t) := \frac{\mbox{\textit{listing-trans}}(u,\Delta t)}{\mbox{\textit{listing-trans}}(u,\Delta t)+\mbox{\textit{pickup-trans}}(u,\Delta t)} ,\] where  
  $\mbox{\textit{listing-trans}}(u,\Delta t)$  is defined as  the number of transactions that contain items listed by user $u$ and occurred in the time interval $\Delta t$, and 
$\mbox{\textit{pickup-trans}}(u,\Delta t)$  is defined as the number of transactions that contain items picked up by user $u$ and occurred in the time interval $\Delta t$.

Lastly, we use a time-series clustering model (see Section~\ref{subsection:Time-series Clustering}) was used to cluster key users into groups with similar behavioral patterns. We utilized the DR measure calculated over time, which represents user behavior, as an input for this model.

To find the most optimal parameters tuning for the clustering model, which will give us meaningful results and prevent minimum distance distortion~\cite{jeong2011weighted}, we tested two parameters. These were: the number of clusters and the distance matrix. Based on previous studies in the field of time-series clustering~\cite{rojas2020estimation, gauthier2021online}, we use the Calinski-Harabasz criterion~\cite{baarsch2012investigation} to determine the optimal number of clusters. In addition, we plot the time-series clustering algorithm results with different distance matrices. Based on this plot, we manually choose one matrix to give meaningful results by unambiguous trends without biases and noises in the trend lines. Our goal is to examine the different trends for the different clusters and prevent minimum distance distortion.

Then, we analyzed the trend lines of key users’ behaviors to identify different patterns of behaviors.

\subsection{Users’ trends prediction}
\label{subsection:Users’ trends prediction}
After separating the key users into groups according to their behaviors, our next goal is to use this information as ground truth, in order to predict the future behavior of every key user. For example, whether a key user will always be a donor, always a recipient, or change his or her behavior from donor to a recipient or vice versa over time. By analyzing the behavior of key users in the first few months after joining the network, we were able to predict future groups. For that, we perform two main steps. We begin feature extraction for each key user includes user features and network features. Then, we utilize these features to construct a prediction model that predicts the group to which the key user will belong in the future. To construct the prediction model, a variety of supervised learning algorithms are utilized and tested against common performance metrics.

These two steps taken to analyze key users’ behavior are described in the following subsections.

\subsubsection{Features Extraction}
\label{subsubsection:Features Extraction}
The first step includes extracting a variety of key user features from the raw data and network structure features. Following previous studies~\cite{kim2010structural,chakraborty2018application}, the network features defined are related to both the individual key user and also to the whole network. Since we are using time-series data, all features must be extracted at the same period from the time a key user started to be active in the network. Therefore, for each key user \textit{u}, we construct \textit{u}’s ego-network \cite{valerioarnaboldi2012analysis}. Namely, for each key user \textit{u}, we define u’s ego network in time \textit{t} as a sub-graph \( G_{u}^{t}: = <V_{u}^{t},E_{u}^{t}>\) with respect to ego node \textit{u} such that  $V_{u}^{t}  := \{ v\in V^{t} \vert   \exists (u,v)\in E^{t}\}$. Edges in the ego network of \textit{u} are denoted by $E_{u}^{t} := \{e=(x,y)\in E^{t} \vert  x=u \lor y \in V_{u}\}$ ~\cite{biswas2015investigating}.

%\forall (u,y)\in E_{u}^{t},y\in V_{u}^{t} \}$ 
Then, for each key user \textit{u}, we extract the following features from it's ego-network graph \(G_{u}^{t}\) in time t:

\begin{itemize}

	\item \textit{Nodes-number(\(G_{u}^{t}\))}  - \( \vert V_{u}^{t}\vert\) -  the number of nodes in the ego-network of key user $u$.

	\item \textit{Edges-number(\(G_{u}^{t}\))} - \( \vert E_{u}^{t}\vert\) -   the number of edges in the ego-network of key user $u$.

	\item \textit{Density(\(G_{u}^{t}\))}~\cite{anderson1999interaction} - density of the network graph of key user $u$ at time $t$. Defined as: \( D^{t}(u) =\frac{\vert E_{u}^{t}\vert }{\vert V_{u}^{t}\vert (\vert V_{u}^{t}\vert -1)}\). 

	\item \textit{PageRank(\(G_{u}^{t}, v\))}~\cite{mihalcea2004pagerank} - In PageRank, each vertex's score is the count of its inbound links. The higher the number is, the higher the importance of the vertex. Moreover, the importance of the vertex determines how important the outbound link is, and this information is also taken into account by the ranking model. Hence, the PageRank score of a vertex \(v\) is determined based on its inbound links and the score of the vertices that connect those links.

	\item \textit{Closeness-centrality(\(G_{u}^{t}, v\))} ~\cite{zhang2017degree} - the number of edges on the shortest path between every two nodes in the graph. if the length of the vertex \(v\) shortest paths with other nodes in the network is small, then this node has a high closeness centrality.

	\item \textit{Clustering-coefficient(\(G_{u}^{t}, v\))}~\cite{saranadivsoffer12005network} - quantifies how well connected a the neighbors of vertex \(v\) are in a graph.
	
	\item\textit{Pickups-count(\(G_{u}^{t}, v\))} - the number of transactions in which the key user represented as vertex \(v\) picked up items until time $t$ (inclusive). Defined as: \( d_{in}^{t}(v)\).
 
    \item\textit{Percent-of-listing-items(\(G_{u}^{t}, v\))} - the number of the listing transactions of a key user, represented as vertex \(v\),  to all of the user transactions at time t. 
    Defined as: \(\frac{d_{out}^{t}(v)}{d_{out}^{t}(v) +d_{in}^{t}(v)}\).

\end{itemize}

In addition to the network features mentioned above, we also use raw features calculated for an initial period of use for each key user. Those features can include \textit{the number of items the key user listed}, \textit{the number of items the user picked up}, \textit{the ratio between the listing items of the key user and the number of the listing and pick-up items}, and \textit{counts of user activity}. Moreover, it is possible to add network-specific features, as we demonstrate in Section~\ref{section:EXPERIMENTS} when analyzing the OLIO network.

\subsubsection{Constructing Prediction Models}
\label{subsubsection:Constructing Prediction Models}
Our method's final step is to use a prediction model based on predefined key user features to predict key user behavior trends. Namely, we generate the prediction models by utilizing the following algorithms: Naïve base~\cite{berrar2018bayes}, Decision tree~\cite{priyank2020decision}, Logistic regression~\cite{walker2007}, Random forest \cite{biau2016random}, Support vector classifier (SVC)~\cite{benhur2001support}, and XGBoost~\cite{chen2015xgboost}. To evaluate the different prediction models’ performances, we use the Accuracy measure and the F1 score~\cite{yacouby2020probabilistic}. Moreover, to better understand which features contribute to users’ classification into a specific group for the algorithm with the highest accuracy in the prediction model, we utilize the SHAP model~\cite{scavuzzo2022feature,antwarg2020explaining} which is based on Shapley values from game theory.

\begin{figure*}[t]
    \centering
    \includegraphics[width=0.8\textwidth]{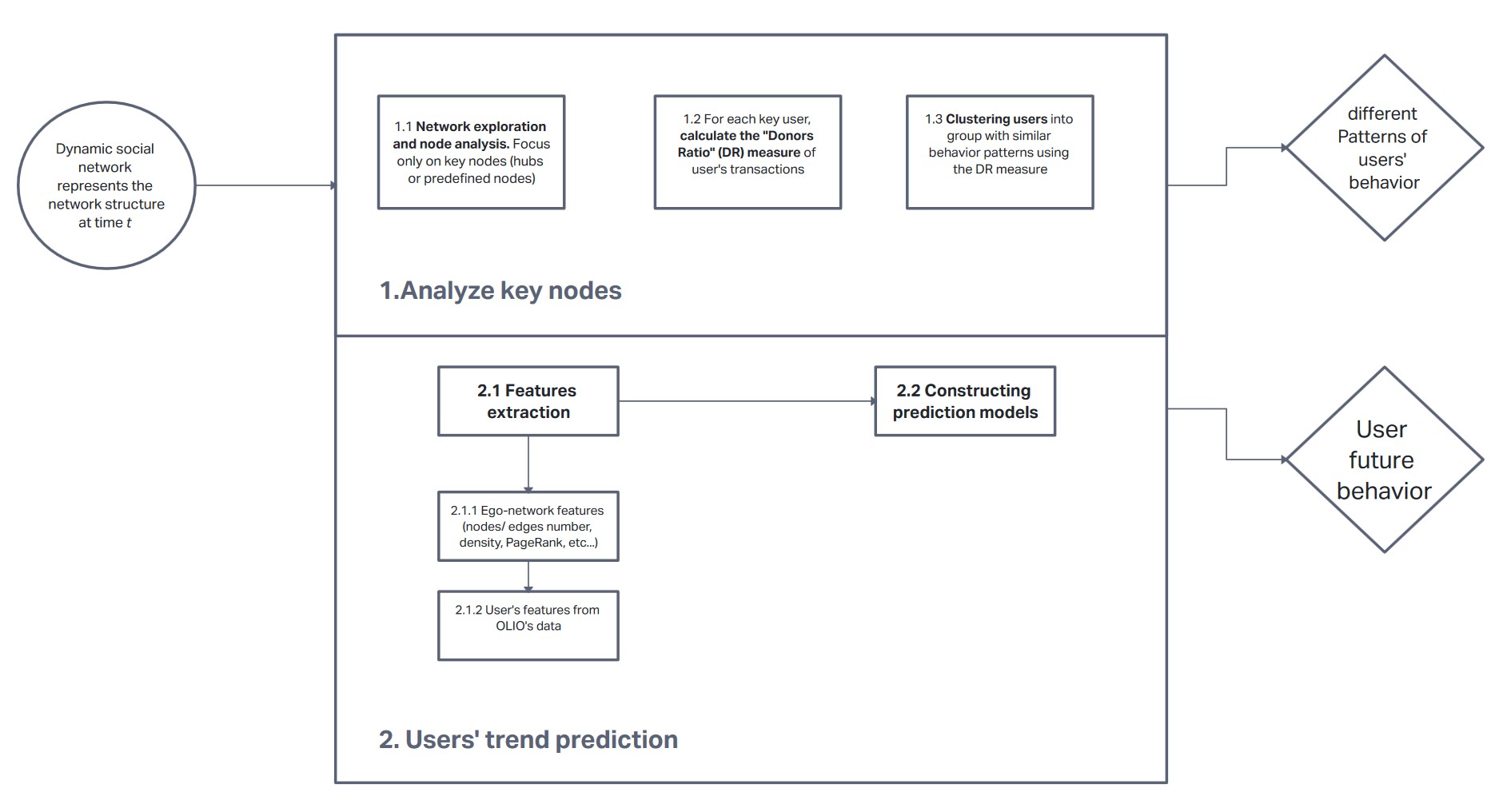}
    \caption{An overview of our methods}
    \label{figure:2}
\end{figure*}

\section{EXPERIMENTS}
\label{section:EXPERIMENTS}
We evaluated our method (see Section~\ref{section:METHODS} and Figure~\ref{figure:2}) on OLIO’S network which contains over 2.48M users and over 2.65M items and specifically on subnetwork (communities) around the UK. Those communities accounted for about 100k users and 300k transactions between 2017 and 2020. The dataset is described in detail in Section~\ref{subsection:Datasets} and the experiment is described in Section~\ref{subsection:Evaluate}.

\subsection{Datasets}
\label{subsection:Datasets}
Our experiment was based on OLIO’s data between April 1$^{st}$, 2017, and July $31^{st}$, 2020. The dataset contained 2,488,673 users and 2,657,683 items worldwide. Our network was structured so that each node represented a user, and each edge represented a transaction between two users. Furthermore, we filtered users who had participated in less than three transactions. 
The filtered network contains 123,602 users and 361,043 transactions.

This study focused only on UK data since the UK has the largest number of active users, which covers 79.7\% of all network users, and also covers 82.8\% of the entire world's transactions (see Table~\ref{table:1}). Additionally, we focused on the largest communities in the UK.

As shown in previous studies (see Section~\ref{subsection:Network analysis}), community detection is a common method for analyzing social network data. By separating the network into communities and analyzing each community separately, we can compare different and smaller groups of users, most likely in different geographical areas. This enabled us to test whether subnetworks act similarly in other locations. 
\begin{figure}[ht]
    \centering
    \captionsetup{justification=centering}
    \includegraphics[width=\linewidth]{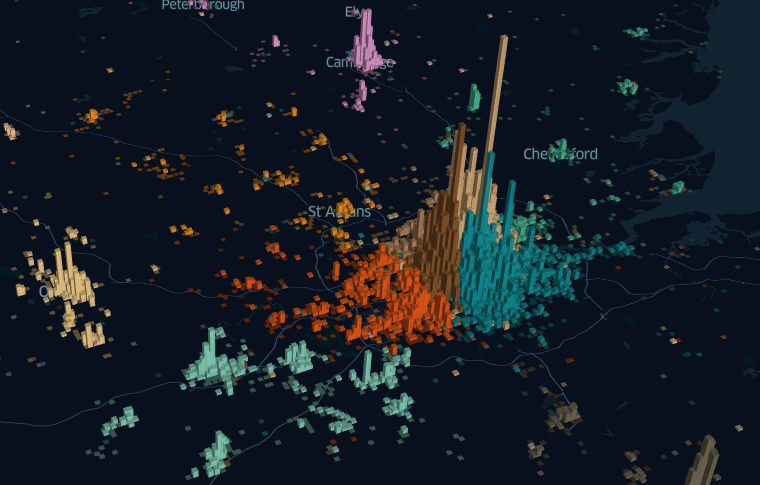}
    \caption{Communities across the United Kingdom. Each color represents a different community. According to this map, we deduce that community detection in the OLIO network is usually carried out according to the geographical distribution of the various users.}
    \label{figure:3}
\end{figure}
%The sizes of both the entire world OLIO network and the size of the United Kingdom's OLIO network are described in Table~\ref{table:1}.
For each community, we analyzed only the first year of activity of \textit{active key users}. Active key users are users who met the following two conditions:

\begin {enumerate}

\item Users whose period between their first and last transaction is at least one year.

\item Users who have at least six weeks in which they listed at least one item each week.\footnote{After analyzing the data of our key users, we concluded that it would be best to concentrate on those who have been active for a minimum of six weeks, as there is a strong correlation between this level of activity and the accuracy of our prediction model.}

\end{enumerate}

Similarly to Bergmeir et al.'s study of validity of cross-validation for evaluating time series prediction~\cite{bergmeir2018note}, we used 10-fold cross validation for evaulating our method.

%Similarly to Ozyegen et al.'s study of time-series classification~\cite{ozyegen2022interpretable}, we split the key users into train and test sets in the separation of 70\% and 30\%, respectively. 

To test and evaluate our second method, the prediction model, we trained our model on the first $t$ months of the key user's activity, as will be explained in detail in the following section. 
%the identification of the users' trends. According to these results, it can be seen that the behavioral trend changes (if there is a change) after 3 months of activity. Therefore, we decided to train data for ($t=3$ months) from the user's first months of activeness and perform the test for his behavior for the rest of the year.
\begin{center}
\begin{table}[!hbt]
\centering
\renewcommand{\arraystretch}{1.3}
\caption{Networks Sizes}
\label{table:1}
\begin{adjustbox}{max width=\linewidth}
\begin{tabular}{ |c|c|c| } 
\hline
 & {\textbf{Nodes (Users)}} & {\textbf{Edges (Transactions)}}\\
\hline
{\textbf{Entire world network}} & 123,602 & 361,043\\
\hline
{\textbf{United Kingdom network}} & 98,521 & 299,104\\
\hline
\end{tabular}
\end{adjustbox}
\end{table}
\end{center}

\subsection{Evaluate the method on real-world OLIO's network}
\label{subsection:Evaluate}
To evaluate our two methods, we created a network from the chosen dataset (see Section~\ref{subsection:Datasets}), and implemented the Louvain community detection algorithm on this network. We performed our methods (see Secion~\ref{section:METHODS}) for the entire network and each community.

Like Makov et al.’s study~\cite{makov2020social} which argued that heroes were the primary source for the platform’s listing, we also found that heroes were the key users in the network and also in each community. We observed that the heroes in most communities had an average degree of 143 more than regular users. Furthermore, although heroes made up a smaller part of the wider communities (about 20\% in the largest network), their transactions constituted a significant part (see Figure~\ref{figure:comparison_users_heroes}) of the transactional activity between the community members (amounting to about 57\% of network transactions in the largest sub-networks). 
Therefore, the focus of this study was on heroes. From the heroes’ list, we filtered out those who did not meet the active users’ criteria~\ref{subsection:Datasets}), and examined only the active members.

To better understand the behavior of the selected heroes, we utilized the DR measure, which examines the ratio of the number of transactions listed by a user to its overall number of transactions within a specific time frame (see Section~\ref{subsection:Analyze key users of the network}). However, since each hero is active for different lengths of time, a standardized measurement process needed to be established. We found that in the largest community, which contains 24\% of the users, only 56\% of heroes were active for more than two years, and just 23.3\% were active for more than three years. The second largest community contained 18.8\% of the users, where 41.1\% of heroes were active for more than two years, and only 8.4\% were active for more than three years. Consequently, we selected the heroes’ first year of activity
(which began with their first transaction) as the time period of analysis, as this is when the majority of heroes engage in transactions.

\begin{figure}[hbt]
    \centering
    \captionsetup{justification=centering}
    \includegraphics[width=\linewidth]{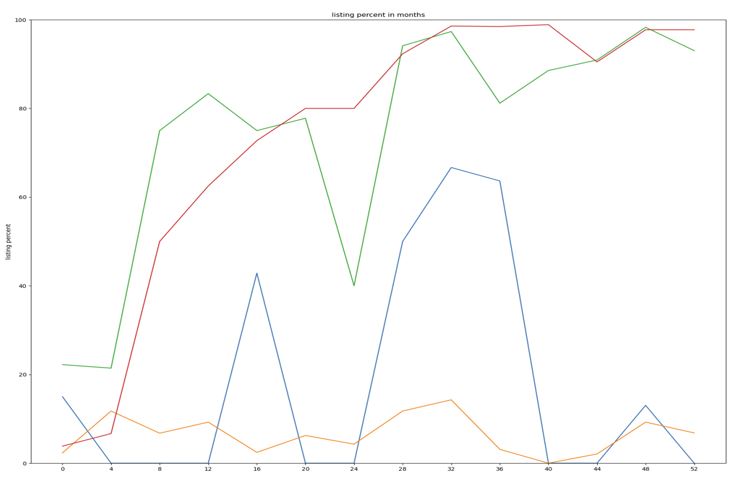}
    \caption{An example DR measure calculation for different users with different trends over time periods of months starting from the first year of the use in OLIO.}
    \label{figure:4}
\end{figure}

The study extracted features for each hero by analyzing their activity during the first t months of their participation in the network. To determine the optimal value of $t$, we analyzed the trend lines of the different clusters resulting from the first algorithm. The value of $t$ represents the minimum number of months needed to identify a change in the behavior pattern of all the clusters. The goal is to identify the specific period within which changes in hero behavior occur if there are any. This is to ensure that this period is consistent across all clusters and helps to optimize the feature extraction process and improves the accuracy of the subsequent machine learning algorithms used to predict future user behavior. The DR measure, which is a way of quantifying behavior, was calculated for a specific period and an example of this is shown in  Figure~\ref{figure:4}.

% Our data refers to a specific time period of about 3 years. Some of the users may have been created before the start date of the dataset (see section~\ref{subsection:Datasets}). In this case, these users will be considered as having started their activity on the start date of the dataset.

In the final step of our first method (see Secion~\ref{subsection:Analyze key users of the network}), we followed the approach proposed by Ruiz et al.~\cite{ruiz2020} and utilized K-means for time-series clustering in our experiments. We conducted a series of tests by applying the Calinski-Harabasz criterion~\cite{baarsch2012investigation} ($\forall k \in [4,10]$) to determine the optimal number of clusters $k$. Additionally, we experimented with three different distance matrices~\cite{ali2019clustering}: Euclidean, Dynamic Time Warping (DTW), and Soft-DTW. In order to select the most suitable distance matrix; we visually examined the K-means results for each of the matrices using the chosen number of clusters and manually chose the best one according to our method.

After choosing the number of clusters and the distance matrix, we plotted the different clusters’ results and analyzed the trend lines to find behavior patterns.

In the second method (see Section~\ref{subsection:Users’ trends prediction}), we constructed a heroes’ trend prediction model based on the results of the experiment of the first algorithm which identified different heroes’ trends. We created a prediction model based on the future behavior of similar users (users in the same groups).

To associate the hero with a specific group, we utilized the heroes’ activity in their $t$ months from the first months of activity in the OLIO application.

In addition to the general network features described in Section~\ref{subsubsection:Features Extraction}, we also extracted features from OLIO's raw data. For each key user, the hero, we calculated the following features:
\begin{itemize}[noitemsep]

\item\textit{Articles Count(\(u,t\))} – the number of items the user \(u\) posted until time $t$ (inclusive).
\item\textit{Messages Count(\(u,t\))} – the number of messages the user \(u\) sent until time $t$ (inclusive).
\item\textit{Rating current(\(u,t\))} – the rating of the user \(u\) until time $t$ (inclusive).
\item\textit{Rating count(\(u,t\))} –  the number of times the user \(u\) was rated until time $t$ (inclusive).
\item\textit{Likes count(\(u,t\))} –  the number of likes made by the user \(u\) until time $t$ (inclusive).
\item\textit{Stories count(\(u,t\))} –  the number of stories posted by the user \(u\) until time $t$ (inclusive).
\item\textit{Comments count(\(u,t\))} –  the number of comments posted by the user \(u\) until time $t$ (inclusive).
\end{itemize}

Combining these features with the network features forms the foundation of our prediction model. It predicts whether key users will always be "active donors", or whether their trend will reverse.

\section{RESULTS}
\label{section:RESULTS}
We evaluated our methods on the entire network in the UK and for the largest communities. By using the Louvain community detection algorithm on the UK OLIO’s network, we uncovered 63 disjoint communities. The most users in the same community were located in the same geographical area (see Figure~\ref{figure:3}), which shows that users mainly interact with other users that are located near them.

%Similar to Makov et al. study~\cite{makov2020social} that argued that heroes were the primary users for listing transactions on the platform, we also found that heroes are the central users in the network and in each community. We observed that the food waste heroes in most communities had an average degree of 143 more than the average degree of the regular users in the community network. In addition, despite the fact that the number of heroes was a small part of the larger communities, the number of their transactions was a significant part of the transactions of the community members (see Figure~\ref{figure:comparison_users_heroes}).
%Therefore, the focus of the project was on food waste heroes.

We analyzed and clustered the heroes by their time-series behavior over one year. According to the Calinski-Harabasz criterion results (see Figure~\ref{figure:calinski_Harabasz_result}), we chose $k=4$ as the number of clusters. For this, we analyzed K-means results with three different distance metrics (see Figure~\ref{figure:10}). According to our method, we manually chose to use the Euclidean distance to be able to examine the different trends for each cluster without minimum distance distortion (see Section~\ref{subsection:Analyze key users of the network}).

By analyzing the behavior line in each cluster (see Figure~\ref{figure:12}), we identified four different types of heroes trend behavior:

\begin {enumerate}

\item \textit{Future Passive Donors (FPD)} - heroes whose initial percentage of listing items is \textit{high and then decreases}.

\item \textit{Stable Active Donors (SAD)} - heroes whose initial percentage of listing items is \textit{high and remains stable}.

\item \textit{Future Active Donors (FAD)} - heroes whose initial percentage of listing items is \textit{low and then increases}.

\item \textit{Stable Passive Donors (SPD)} - heroes whose initial percentage of listing items is\textit{ low and remains stable}. 

\end{enumerate}

%For the largest community, FPD group contains 22.18\% of the users, SAD group contains 30.11\%, FAD group contains 25.18\%, and SPD group contains 22.54\%.

%As we can see in Figure~\ref{figure:11}, users in different clusters are located in various and different locations around the UK.

We found two prominent cases of heroes behavior: 

\begin{itemize}

\item\textit{"Starting high" case - groups FPD+SAD} – groups of users whose initial listing items percentage is high (represents the donors).
\item\textit{"Starting low" case - groups FAD+SPD} – groups of users whose initial listing items percentage is low (represents the recipients - passive donors).

\end{itemize}

\begin{figure}
    \centering
    \captionsetup{justification=centering}
    \includegraphics[width=0.8\linewidth]{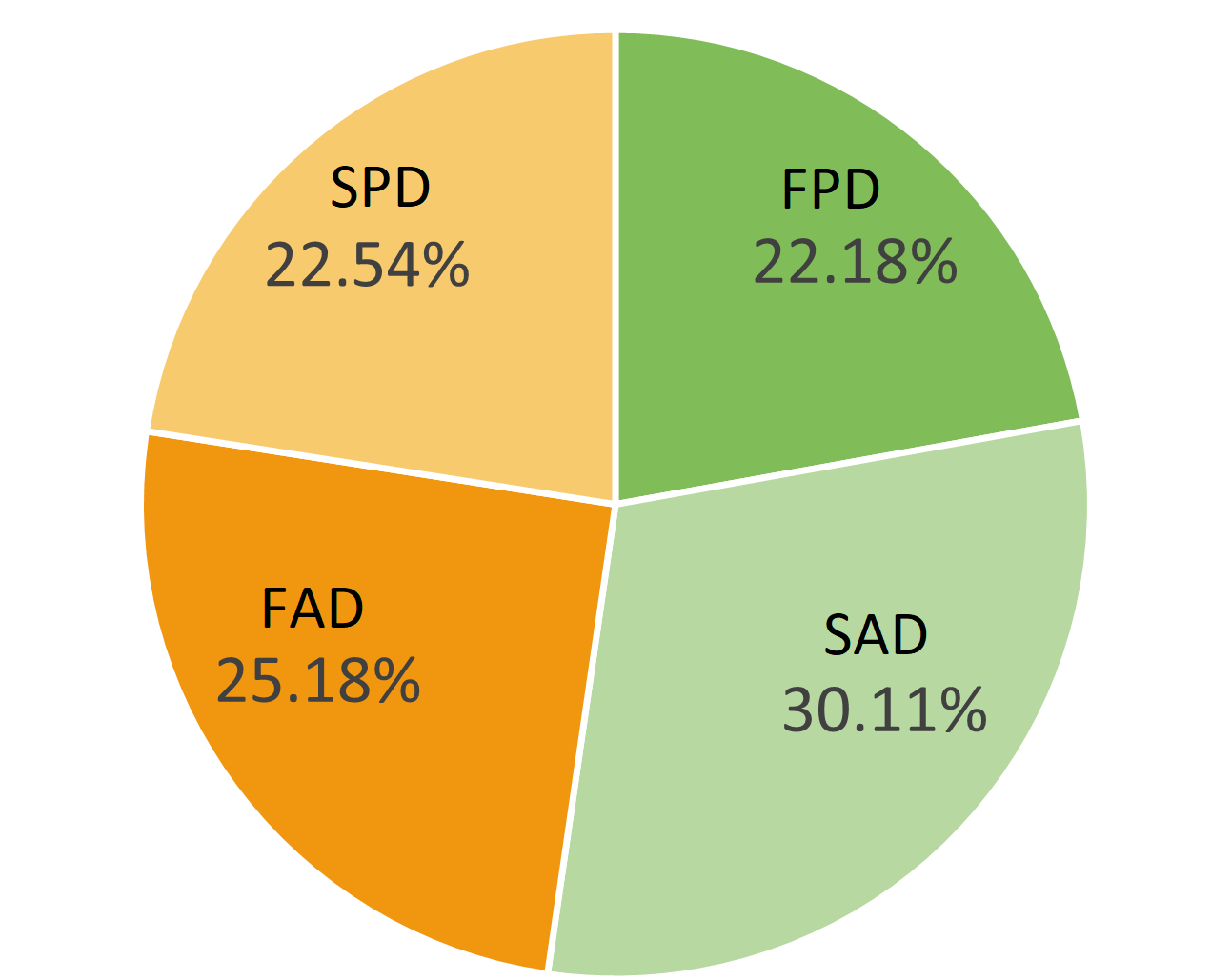}
    \caption{Users percentage in each group in the largest community}
    \label{figure:groups_pie_chart}
\end{figure}

To test and evaluate our second method for key user trend prediction, we examined the above cases resulting from the first method. We tested each case for the entire UK network, and for the two largest communities in the UK network.

In our approach, we linked each tested hero with one of the behavior groups to make predictions about their future behavior. This allows us to analyze the two behavior trends of a particular group and use that information to make informed predictions about an individual hero’s future actions. For example, if a hero is identified as belonging to the FAD group, we can predict that in the future they will become a passive donor. This means that the percentage of items they have listed will decrease over time, based on the behavior patterns observed within the FAD group. By associating each hero with a specific group, we can draw insights about how their behavior may change over time and use that information to generate our predictions.

According to Figure~\ref{figure:12}, which resulted from the first method, it is evident that the behavioral trend changes (if there is a change) after about three months of activity. Therefore, we decided to choose ($t=3$ months), extract features for the user's ($t=3$ months) of activity, and perform the test for the remaining part of the year.

Table~\ref{table:2} presents the number of heroes and transactions in each network. Tables 3 and 4 present the results predicted by different prediction models for the entire UK network and the two largest communities in this network. For the entire UK network, our method utilizing the XGBoost algorithm obtained the highest accuracy score (79.1\%) for the "starting high" case. For the largest community, in the same case, our method using the SVC algorithm obtained the highest accuracy score (84.6\%).

\begin{figure}
    \centering
    \captionsetup{justification=centering}
    \includegraphics[width=\linewidth]{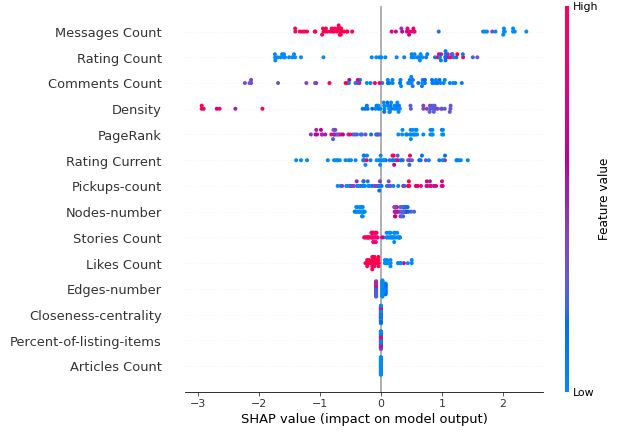}
    \caption{The prediction model features are arranged from top to bottom according to their importance to the outcome of the XGBoost model. These results were calculated through the SHAP method. Each point represents a user feature value. \textit{Messages count} is the feature with the highest impact on the model. Users with high values of \textit{Messages count} will belong, with higher probability, to the SPD group, while users with low values will belong, with higher probability, to the FAD group. In addition, we can observe that \textit{Closeness centrality}, is not an important feature and has almost no contribution to the prediction, whether its values are high or low.}
    \label{figure:shap}
\end{figure}

\begin{figure*}
    \centering
    \captionsetup{justification=centering}
    \includegraphics[width=\linewidth]{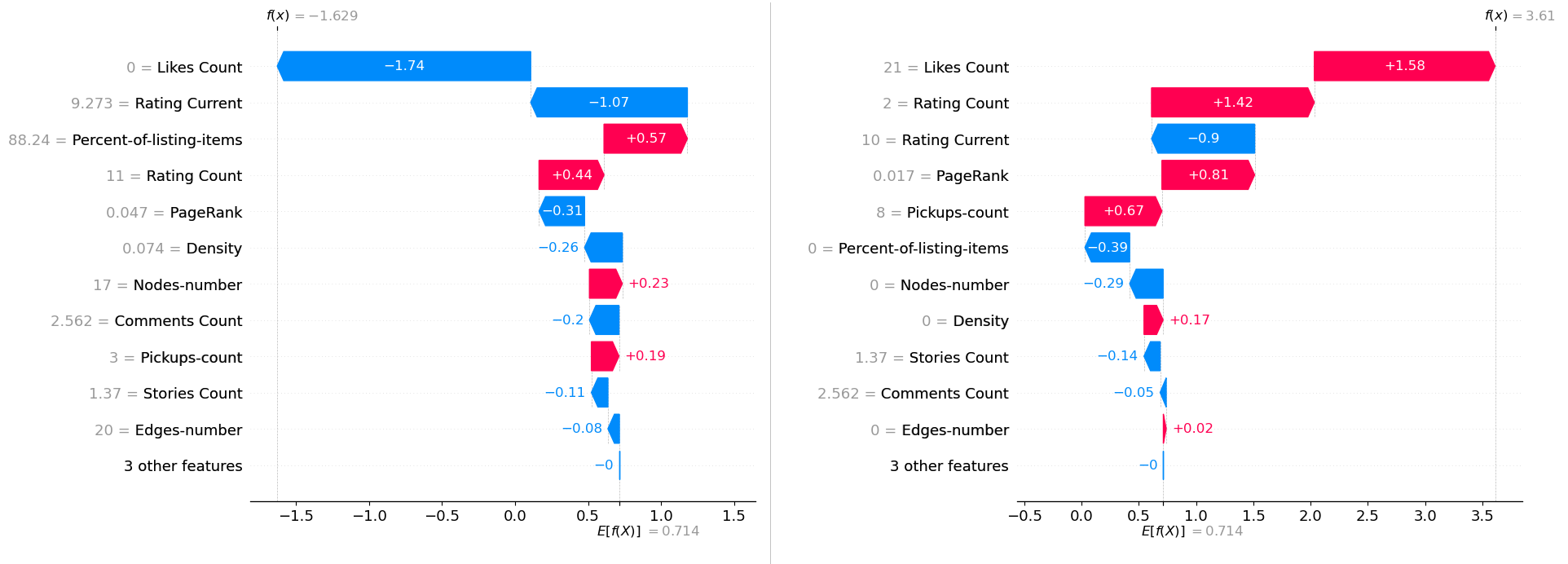}
    \caption{This figure presents an analysis of two different users who is their initial state is active donors. This analysis is based on our prediction model result and its feature importance analysis using SHAP. The user in the left diagram changes his behavior to a passive donor over time. The user in the right diagram does not change his or her behavior and stays an active donor. We present the features which contributed the most to these specific users regarding their behavior over time. For the left user, the most significant features were \textit{Likes count} and \textit{Rating current}. The gray text before the feature names shows the value of each feature for this sample. The value 9.273 of \textit{Rating current} has a negative contribution to the model and a passive donor future prediction. For the right user, the most significant features were \textit{Likes count} and \textit{Rating count}. The value 21 of "Likes count" has a positive contribution to the model and an active donor future prediction.}
    \label{figure:two_users_starting_active}
\end{figure*}

In addition, as explained in the method, we calculated SHAP values for the prediction model with the highest accuracy: XGBoost. Figures~\ref{figure:shap} and ~\ref{figure:two_users_starting_active} present insight into how the contribution of an individual feature to the model output is affected by those values. The features \textit{Message Count}, \textit{Rating Count}, and \textit{Comment Count} were the most significant features in determining similar users.

%\begin{figure}[H]
 %   \centering
  %  \captionsetup{justification=centering}
   % \includegraphics[width=0.9\linewidth]{Fig11.JPG}
    %\caption{Users in different clusters across the United Kingdom. 
    %Each circle represents a user and each color represents a different cluster from the identified cluster in the first method. Namely, each cluster contains users from different areas in the United Kingdom.}
    %\label{figure:11}
%\end{figure}

\begin{table*}
\centering
\renewcommand{\arraystretch}{1.3}
\captionsetup{justification=centering}
\caption{The number of users and transactions for each tested network}
\label{table:2}
\begin{adjustbox}{max width=\textwidth}
\begin{tabular}{p{2.88cm}p{2.54cm}p{2.38cm}p{2.46cm}p{2.43cm}p{2.46cm}p{2.65cm}}
\hline
\multicolumn{1}{|p{2.35cm}}{} & 
\multicolumn{2}{|p{3.86cm}}{\textbf{All network}} & 
\multicolumn{2}{|p{3.84cm}}{\textbf{\#1 largest community}} & 
\multicolumn{2}{|p{4.05cm}|}{\textbf{\#2 largest community}}\\ 
\hline
\multicolumn{1}{|p{2.35cm}}{} & 
\multicolumn{1}{|p{2.01cm}}{\textbf{"Starting high" case}} & 
\multicolumn{1}{|p{1.85cm}}{\textbf{"Starting low" case}} & 
\multicolumn{1}{|p{1.93cm}}{\textbf{"Starting high" case}} & 
\multicolumn{1}{|p{1.9cm}}{\textbf{"Starting low" case}} & 
\multicolumn{1}{|p{1.93cm}}{\textbf{"Starting high" case}} & 
\multicolumn{1}{|p{2.12cm}|}{\textbf{"Starting low" case}}\\ 
\hline
\multicolumn{1}{|p{2.35cm}}{\textbf{Number of heroes}} & 
\multicolumn{1}{|p{2.01cm}}{271} & 
\multicolumn{1}{|p{1.85cm}}{298} & 
\multicolumn{1}{|p{1.93cm}}{70} & 
\multicolumn{1}{|p{1.9cm}}{67} & 
\multicolumn{1}{|p{1.93cm}}{58} & 
\multicolumn{1}{|p{2.12cm}|}{49}\\ 
\hline
\multicolumn{1}{|p{2.35cm}}{\textbf{Number of transactions}} & 
\multicolumn{1}{|p{2.01cm}}{857,124} & 
\multicolumn{1}{|p{1.85cm}}{392,955} & 
\multicolumn{1}{|p{1.93cm}}{115,747} & 
\multicolumn{1}{|p{1.9cm}}{63,815} & 
\multicolumn{1}{|p{1.93cm}}{68,681} & 
\multicolumn{1}{|p{2.12cm}|}{62,777} \\ 
\hline
\end{tabular}
\end{adjustbox}
\vspace*{1 cm}
%\end{table*}
%\begin{table*}
\renewcommand{\arraystretch}{1.3}
\captionsetup{justification=centering}
\caption{Results of our second method with different prediction models for the "Starting high" case and the "Starting low" case tested on the entire network}
\label{table:3}
\begin{adjustbox}{max width=\textwidth}
\begin{tabular}{p{4.6cm}p{3.33cm}p{3.36cm}p{3.23cm}p{3.28cm}}
\hline
\multicolumn{1}{|p{4.6cm}}{\multirow{2}{*}{\parbox{4.6cm}{\textbf{Prediction model} \newline
}}} & 
\multicolumn{2}{|p{6.69cm}}{\textbf{"Starting high" case}} & 
\multicolumn{2}{|p{6.51cm}|}{\textbf{"Starting low" case}} \\ 
\hhline{~----}
\multicolumn{1}{|p{4.6cm}}{} & 
\multicolumn{1}{|p{3.33cm}}{\textbf{Accuracy}} & 
\multicolumn{1}{|p{3.36cm}}{\textbf{F1 score}} & 
\multicolumn{1}{|p{3.23cm}}{\textbf{Accuracy}} & 
\multicolumn{1}{|p{3.28cm}|}{\textbf{F1 score}} \\ 
\hline
\multicolumn{1}{|p{4.6cm}}{\textbf{Naïve base}} & 
\multicolumn{1}{|p{3.33cm}}{0.603} & 
\multicolumn{1}{|p{3.36cm}}{0.237} & 
\multicolumn{1}{|p{3.23cm}}{0.411} & 
\multicolumn{1}{|p{3.28cm}|}{0.36} \\ 
\hline
\multicolumn{1}{|p{4.6cm}}{\textbf{Decision tree}} & 
\multicolumn{1}{|p{3.33cm}}{0.712} & 
\multicolumn{1}{|p{3.36cm}}{0.664} & 
\multicolumn{1}{|p{3.23cm}}{0.568} & 
\multicolumn{1}{|p{3.28cm}|}{\textbf{0.423}} \\ 
\hline
\multicolumn{1}{|p{4.6cm}}{\textbf{Logistic regression}} & 
\multicolumn{1}{|p{3.33cm}}{0.734} & 
\multicolumn{1}{|p{3.36cm}}{0.713} & 
\multicolumn{1}{|p{3.23cm}}{0.621} & 
\multicolumn{1}{|p{3.28cm}|}{0.053} \\ 
\hline
\multicolumn{1}{|p{4.6cm}}{\textbf{Random forest}} & 
\multicolumn{1}{|p{3.33cm}}{0.788} & 
\multicolumn{1}{|p{3.36cm}}{0.751} & 
\multicolumn{1}{|p{3.23cm}}{0.542} & 
\multicolumn{1}{|p{3.28cm}|}{0.212} \\ 
\hline
\multicolumn{1}{|p{4.6cm}}{\textbf{SVC}} & 
\multicolumn{1}{|p{3.33cm}}{0.557} & 
\multicolumn{1}{|p{3.36cm}}{0.101} & 
\multicolumn{1}{|p{3.23cm}}{\textbf{0.632}} & 
\multicolumn{1}{|p{3.28cm}|}{0.0} \\ 
\hline
\multicolumn{1}{|p{4.6cm}}{\textbf{XGBoost}} & 
\multicolumn{1}{|p{3.33cm}}{\textbf{0.791}} & 
\multicolumn{1}{|p{3.36cm}}{\textbf{0.778}} & 
\multicolumn{1}{|p{3.23cm}}{0.56} & 
\multicolumn{1}{|p{3.28cm}|}{0.404} \\ 
\hline
\end{tabular}
\end{adjustbox}
%\end{table*}
\vspace*{1 cm}
%\begin{table*}
\centering
\renewcommand{\arraystretch}{1.3}
\captionsetup{justification=centering}
\caption{Results of our second method with different prediction models for the "Starting high" case and the "Starting low" case tested on the two largest communities (by users) in the United Kingdom}
\label{table:4}
\begin{adjustbox}{max width=\textwidth}
\begin{tabular}{p{2.14cm}p{1.8cm}p{1.3cm}p{1.83cm}p{1.27cm}p{1.83cm}p{1.24cm}p{1.83cm}p{1.24cm}p{1.83cm}p{1.27cm}p{1.72cm}p{1.24cm}}
\hline
\multicolumn{1}{|p{2.14cm}}{\multirow{3}{*}{\parbox{2.14cm}{\textbf{Prediction model}}}} & 
\multicolumn{4}{|p{6.19cm}}{\textbf{\#1 largest community}} & 
\multicolumn{4}{|p{6.14cm}|}{\textbf{\#2 largest community}} \\ 
\hhline{~------------}
\multicolumn{1}{|p{2.14cm}}{} & 
\multicolumn{2}{|p{3.1cm}}{\textbf{"Starting high" case}} & 
\multicolumn{2}{|p{3.1cm}}{\textbf{"Starting low" case}} & 
\multicolumn{2}{|p{3.07cm}}{\textbf{"Starting high" case}} & 
\multicolumn{2}{|p{3.07cm}|}{\textbf{"Starting low" case}} \\ 
\hhline{~------------}
\multicolumn{1}{|p{2.14cm}}{} & 
\multicolumn{1}{|p{1.8cm}}{\textbf{Accuracy}} & 
\multicolumn{1}{|p{1.3cm}}{\textbf{F1 score}} & 
\multicolumn{1}{|p{1.83cm}}{\textbf{Accuracy}} & 
\multicolumn{1}{|p{1.27cm}}{\textbf{F1 score}} & 
\multicolumn{1}{|p{1.83cm}}{\textbf{Accuracy}} & 
\multicolumn{1}{|p{1.24cm}}{\textbf{F1 score}} & 
\multicolumn{1}{|p{1.83cm}}{\textbf{Accuracy}} & 
\multicolumn{1}{|p{1.24cm}|}{\textbf{F1 score}}\\ 
\hline
\multicolumn{1}{|p{2.14cm}}{\textbf{Naïve base}} & 
\multicolumn{1}{|p{1.8cm}}{0.413} & 
\multicolumn{1}{|p{1.3cm}}{\textbf{0.333}} & 
\multicolumn{1}{|p{1.83cm}}{0.757} & 
\multicolumn{1}{|p{1.27cm}}{0.423} & 
\multicolumn{1}{|p{1.83cm}}{0.602} & 
\multicolumn{1}{|p{1.24cm}}{\textbf{0.658}} & 
\multicolumn{1}{|p{1.83cm}}{0.654} & 
\multicolumn{1}{|p{1.24cm}|}{0.532} \\ 
\hline
\multicolumn{1}{|p{2.14cm}}{\textbf{Decision tree}} & 
\multicolumn{1}{|p{1.8cm}}{0.581} & 
\multicolumn{1}{|p{1.3cm}}{0.167} & 
\multicolumn{1}{|p{1.83cm}}{0.843} & 
\multicolumn{1}{|p{1.27cm}}{\textbf{0.663}} & 
\multicolumn{1}{|p{1.83cm}}{0.521} & 
\multicolumn{1}{|p{1.24cm}}{0.54} & 
\multicolumn{1}{|p{1.83cm}}{0.58} & 
\multicolumn{1}{|p{1.24cm}|}{0.567}\\ 
\hline
\multicolumn{1}{|p{2.14cm}}{\textbf{Logistic regression}} & 
\multicolumn{1}{|p{1.8cm}}{0.625} & 
\multicolumn{1}{|p{1.3cm}}{0.16} & 
\multicolumn{1}{|p{1.83cm}}{0.625} & 
\multicolumn{1}{|p{1.27cm}}{0.0} & 
\multicolumn{1}{|p{1.83cm}}{0.54} & 
\multicolumn{1}{|p{1.24cm}}{0.542} & 
\multicolumn{1}{|p{1.83cm}}{0.36} & 
\multicolumn{1}{|p{1.24cm}|}{0.3} \\ 
\hline
\multicolumn{1}{|p{2.14cm}}{\textbf{Random forest}} & 
\multicolumn{1}{|p{1.8cm}}{0.754} & 
\multicolumn{1}{|p{1.3cm}}{0.0} & 
\multicolumn{1}{|p{1.83cm}}{0.841} & 
\multicolumn{1}{|p{1.27cm}}{0.44} & 
\multicolumn{1}{|p{1.83cm}}{0.503} & 
\multicolumn{1}{|p{1.24cm}}{0.37} & 
\multicolumn{1}{|p{1.83cm}}{0.524} & 
\multicolumn{1}{|p{1.24cm}|}{0.628} \\ 
\hline
\multicolumn{1}{|p{2.14cm}}{\textbf{SVC}} & 
\multicolumn{1}{|p{1.8cm}}{\textbf{0.846}} & 
\multicolumn{1}{|p{1.3cm}}{0.0} & 
\multicolumn{1}{|p{1.83cm}}{0.77} & 
\multicolumn{1}{|p{1.27cm}}{0.0} & 
\multicolumn{1}{|p{1.83cm}}{0.42} & 
\multicolumn{1}{|p{1.24cm}}{0.0} & 
\multicolumn{1}{|p{1.83cm}}{0.55} & 
\multicolumn{1}{|p{1.24cm}|}{0.6} \\ 
\hline
\multicolumn{1}{|p{2.14cm}}{\textbf{XGBoost}} & 
\multicolumn{1}{|p{1.8cm}}{0.602} & 
\multicolumn{1}{|p{1.3cm}}{0.0} & 
\multicolumn{1}{|p{1.83cm}}{\textbf{0.896}} & 
\multicolumn{1}{|p{1.27cm}}{0.657} & 
\multicolumn{1}{|p{1.83cm}}{\textbf{0.659}} & 
\multicolumn{1}{|p{1.24cm}}{0.561} & 
\multicolumn{1}{|p{1.83cm}}{\textbf{0.722}} & 
\multicolumn{1}{|p{1.24cm}|}{\textbf{0.709}}\\ 
\hline
\end{tabular}
\end{adjustbox}
\end{table*}

\section{DISCUSSION}
\label{section:DISCUSSION}
After developing our method and conducting experiments to analyze the results, we arrived at the following conclusions. First, our proposed algorithm can be applied to any volunteer-based network, as demonstrated through its successful implementation within OLIO’s network. The features used in our prediction models are mostly generic, and can be applied to any network, including edge number, density, and PageRank. Therefore, it is possible to test our algorithm on various transaction-based networks so as to identify and predict the behavior patterns of key users within those networks in a number of contexts.

%Second, we suggested two different ways to identify the key users, a predefined list of users or using the hub definition. These two options are compatible with the generic algorithm described before. In our experiment, we proved that the predefined key users, the heroes, perform the largest percentage of transactions among users and are indeed the key users.

%By analyzing the results presented in Section~\ref{section:RESULTS}, the following can be noted:

Second, we used the Calinski-Harabas criteria and tested ($\forall k \in [4,10]$). For these $k$ values, we found that $k=4$ is the optimal number of groups. It is possible to test the proposed method with various $k$ values to determine whether selecting a different number of groups for the time-series clustering algorithm will result in detecting additional behavior patterns. 

Third, our method defines a novel behavior measure (DR). While testing this method on the OLIO network, we found two behaviors arising from the DR measure: donors who have a high listing percentage of transactions and recipients who have a low listing percentage of transactions. Some users stay stable during their first year of activity whilst some change their behavior. We also discovered that we could predict users’ future volunteering patterns in OLIO by analyzing their behavioral patterns in the time interval of the first three months from their joining the network. In different networks, the time interval for training of the ML prediction models will need to be adjusted according to the behavioral patterns of the network.

Fourth, our method was tested on the OLIO network and has yielded high accuracy ratings across the UK as well as in its two largest communities. In particular, the classifiers that attained the highest scores are SVC (up to 85.7\% accuracy) and XGBoost (up to 89.6\% accuracy). It is important to determine the most effective classifier for each individual network. Further research could investigate the reasons why one classifier may perform better than another in a specific network.

Fifth, we analyzed data from different locations around the UK. We observed that most users’ transactions were carried out in relatively close geographic areas. For that reason, the communities created based on the transactions are divided mainly by geographical regions (see Figure~\ref{figure:3}). In addition, after analyzing the locations of the users clustered to each behavior group, we observed that users with the same behavior pattern are located in different locations across the UK. Namely, all patterns found in OLIO (FPD, SAD, FAD, and SPD) are, in many cases, generic to different geographical areas in the UK and are not to a specific geographic location.

Sixth, our model performed better on the largest sub-network (see table~\ref{table:4}) in the UK compared to the entire UK (see table~\ref{table:3}) network, particularly in the "starting low" scenario. This leads us to believe that our method is more suitable for strongly connected graphs, such as the largest sub-networks or communities in the UK. As a result, our future research will focus on testing our method specifically on these types of networks, which will be generated using a community detection algorithm.

Lastly, to find in advanced future active donors or passive active donors we need to understand which features are more significant and contribute to the future behavior. We found that the most significant features according to SHAP values are \textit{Message Count}, \textit{Rating Count}, and \textit{Comments Count}. These features are not related to a specific group or the relations among users but to the users’ data. According to the SHAP analysis, these features highly impact the model prediction. For example, \textit{Messages count} is the feature with the highest impact on the model. Heroes with high values of the \textit{Messages count} will belong, in high probability, to the SPD group, while users with low values will belong, in high probability, to the FAD group. In addition, we observed that some of the tested features (such as closeness centrality) have negligible contributions to the prediction model, regardless of whether their values are high or low.

\section{CONCLUSIONS AND FUTURE WORK}
\label{section:CONCLUSION}
This study analyzed users’ behavior in volunteer-based networks, where we suggested two new methods. The first being the analysis of key users and the identification of users’ behavior trends. We defined a new metric for measuring the users’ behavior and then clustered the key users according to this measure. The second method predicted user behavior (see Section~\ref{section:METHODS}).In the second method, we extracted user features from raw data and network features from the user’s ego network, to construct a prediction model for the user behavior trend.

We tested our method on OLIO’s network, which aims to reduce global food waste. In this study, we focused only on data from the UK. Using our first method on OLIO’s network, we managed to identify four different user behaviors. Utilizing the XGBoost model, we were able to predict future user behavior with up to 90.5\% accuracy using our second method.

There are many potential avenues for future research.
Our method can be used immediately to gain insights into any data modeled as a volunteer-based network. There is also scope to test OLIO’s data in multiple geographical locations, building on our findings from UK data. Furthermore, future studies can include socio-geographic and economic features related to tested users or the living neighborhood. Other features that warrant future attention include the number and price rate of supermarkets in the community’s neighborhood. As mentioned in section~\ref{section:DISCUSSION}, we can test different ranges of values for the Calinski-Harabas criteria to find the optimal number of clusters. In addition, it is possible to test other methods for choosing this optimal number as Silhouette~\cite{maharaj2019time}. Also, we can try different time-series clustering methods as k-medoids~\cite{maharaj2019time}, or methods in which the number of clusters does not have to be specified in advance, i.e., the “Snob” clustering method~\cite{bandara2020forecasting}.

\section{ACKNOWLEDGMENTS}
\label{section:Acknowledgments}
We thank OLIO for providing the data for this study. We thank Polly Hember for proofreading this article. In addition, while drafting this article, we used ChatGPT for slight editing according to necessity.

\bibliographystyle{unsrt}
\bibliography{REFERENCES}
\newpage
\appendix
\counterwithin{figure}{section}
\section{APPENDIX}

\begin{figure*}[ht]
    \centering
    \includegraphics[width=0.9\linewidth]{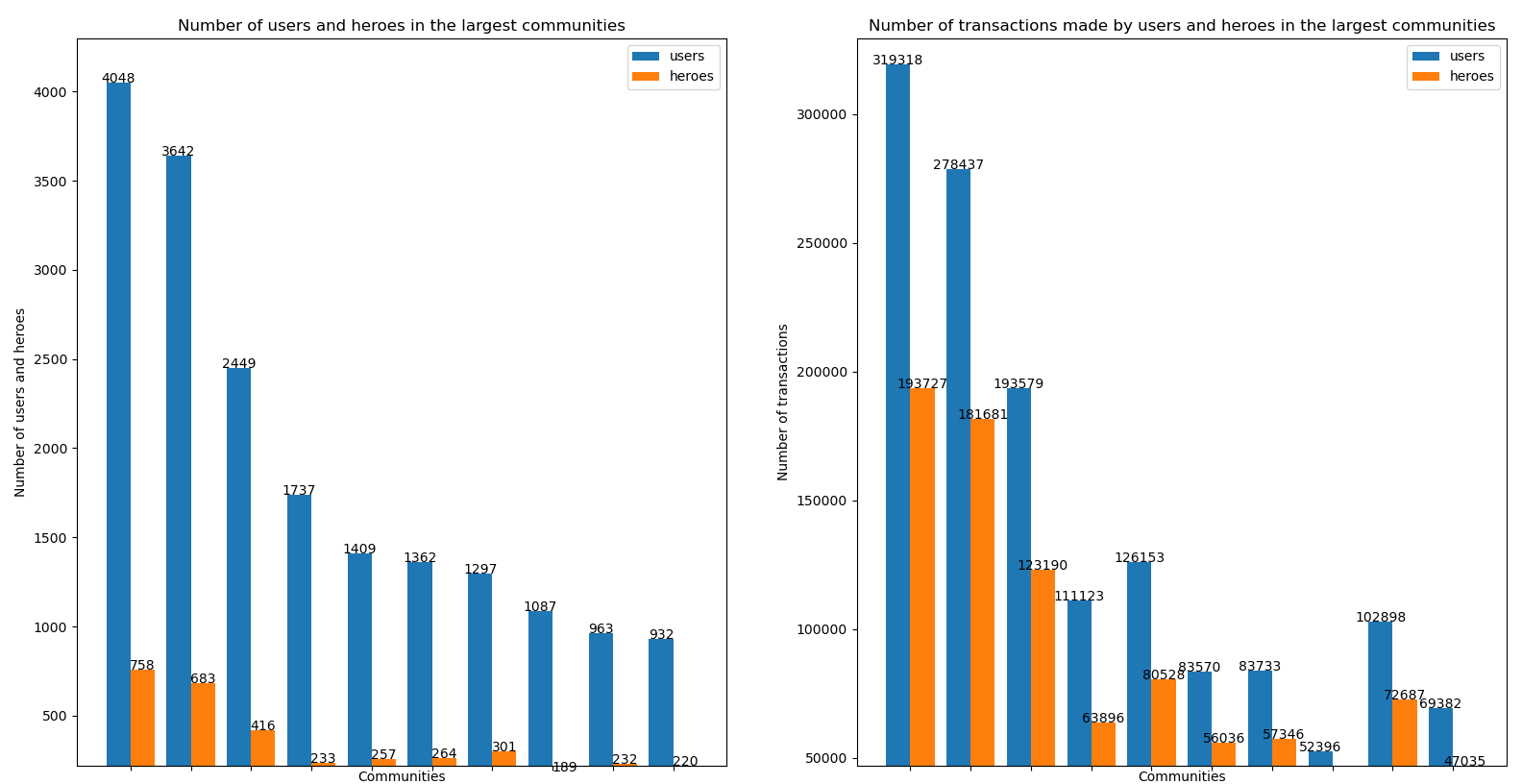}
    \captionsetup{justification=centering}
    \caption{Comparison between regular users and heroes in the largest communities. The x-axis represents the different communities. The plots are sorted descending by the communities' sizes. Each bar in the left plot represents the same community as the bar in the same location in the right plot.}
    \label{figure:comparison_users_heroes}
\end{figure*}

\begin{figure*}[ht]
    \centering
    \includegraphics[width=0.7\linewidth]{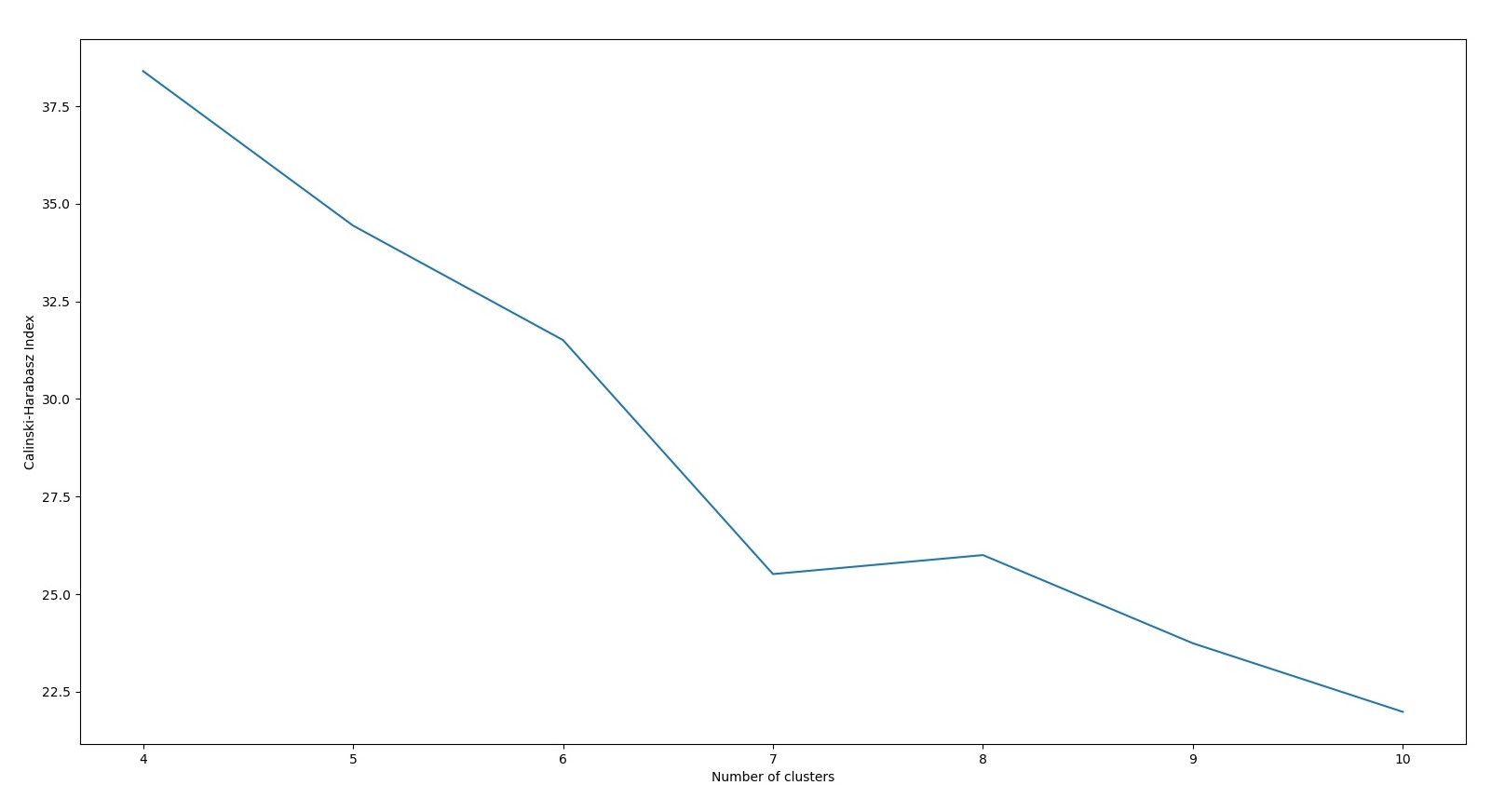}
    \caption{The Calinski-Harabas results for \(k\in[4,10]\)}
    \label{figure:calinski_Harabasz_result}
\end{figure*}

\begin{figure*}[t!]
    \centering
    \includegraphics[width=0.9\linewidth]{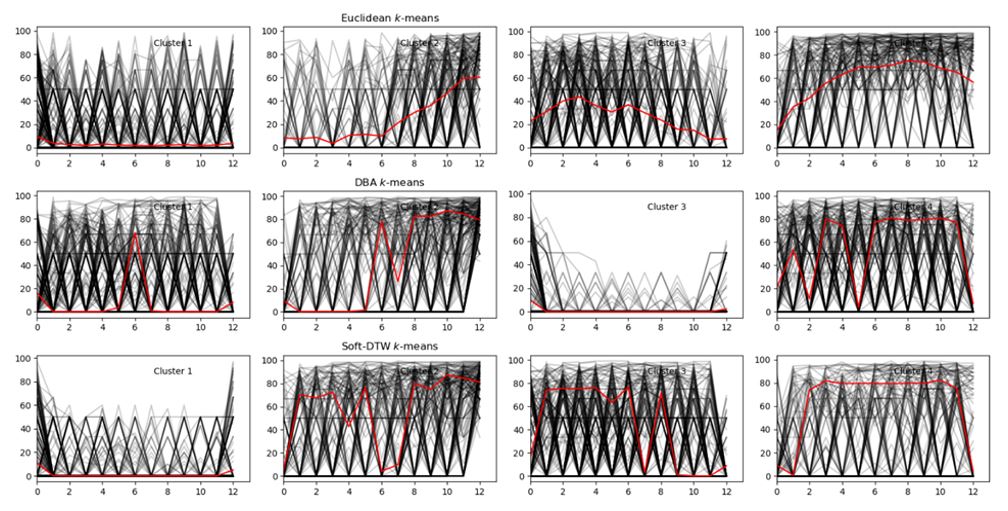}
    \caption{Time-series K-means result with three distance metrics for K=4}
    \label{figure:10}
\end{figure*}

\end{document}